\documentclass[
aip,
amsmath,
amssymb,
reprint,
]{revtex4-1}

\usepackage{graphicx}
\usepackage{dcolumn}
\usepackage{bm}

\usepackage[utf8]{inputenc}
\usepackage[T1]{fontenc}
\usepackage{mathptmx}

\usepackage{siunitx}
\usepackage{stackrel}

\begin{document}


\title[Spin-Wave Optical Elements: Towards Spin-Wave Fourier Optics]{Spin-Wave Optical Elements: Towards Spin-Wave Fourier Optics}

\author{M. Vogel}
\email{mvogel@physik.uni-kl.de}
\affiliation{Department of Physics and State Research Center OPTIMAS,  Technische Universität Kaiserslautern (TUK), Erwin-Schrödinger-Str. 56, 67663 Kaiserslautern, Germany}

\author{B. Hillebrands}
\affiliation{Department of Physics and State Research Center OPTIMAS,  Technische Universität Kaiserslautern (TUK), Erwin-Schrödinger-Str. 56, 67663 Kaiserslautern, Germany}

\author{G. von Freymann}
\affiliation{Department of Physics and State Research Center OPTIMAS,  Technische Universität Kaiserslautern (TUK), Erwin-Schrödinger-Str. 56, 67663 Kaiserslautern, Germany}
\affiliation{Fraunhofer-Institute for Industrial Mathematics ITWM, Fraunhofer-Platz 1, 67663 Kaiserslautern, Germany}

\date{\today}

\begin{abstract}
We perform micromagnetic simulations to investigate the propagation of spin-wave beams through spin-wave optical elements. Despite spin-wave propagation in magnetic media being strongly anisotropic, we use axicons to excite spin-wave Bessel-Gaussian beams and gradient-index lenses to focus spin waves in analogy to conventional optics with light in isotropic media. Moreover, we demonstrate spin-wave Fourier optics using gradient-index lenses. These results contribute to the growing field of spin-wave optics.
\end{abstract}

\maketitle

Spin waves are eigenmodes of a magnetic solid body -- magnons are their quasiparticles. Spin-wave dispersion relations and the corresponding magnonic refractive indices depend not only on the orientation of the external magnetic field but also on several other parameters. Hence, spin-wave propagation is very complex and strongly anisotropic in contrast to light in isotropic media. Recently, spin-wave optics gains increasing attention as it can be seen as a prerequisite for spin-wave based data-processing. In 2016, Gruszecki \textit{et~al.} showed theoretically how to excite Gaussian spin-wave beams using narrowed microwave antennas~\cite{Gruszecki2016}. In the same year, Snell's law of refraction has been verified experimentally in a magnonic system~\cite{Stigloher2016}. Also in 2016, the concept of spin-wave fibers to guide magnons using the \mbox{Dzyaloshinskii}-\mbox{Moriya} interaction has been shown theoretically by Xing \textit{et~al.}~\cite{Xing2016} and Yu \textit{et~al.}~\cite{Yu2016} independently. Moreover, different concepts to focus spin waves using the modulation of the film thickness\cite{Toedt2016} or heating via a laser spot~\cite{Dzyapko2016} have been introduced. Further spin-wave optical elements have been developed in the following year such as Fresnel~\cite{Graefe2017} and Luneburg lenses~\cite{Whitehead2018}. The concept of spin-wave beam excitation via narrowed antennas has been demonstrated experimentally by Körner \textit{et~al.} in the same year~\cite{Koerner2017}. In addition to the described concepts, gradients of the magnetic field and the saturation magnetization have the potential to guide spin waves in magnetic media~\cite{Davies2015_a, Davies2015_b, Gruszecki2018, Vogel2018}. Moreover, magnonic crystals~\cite{Krawczyk2014,Vogel2015,Chumak2017} have the potential to control and manipulate spin-wave propagation via tailored dispersion relations as well.

In this Letter we demonstrate that the powerful concept of Fourier optics not only holds for out-of-plane magnetized magnetic films with an in-plane isotropic spin-wave dispersion relation but also for in-plane magnetization and its corresponding anisotropic spin-wave dispersion.
In order to obtain a deeper insight we perform micromagnetic simulations. Using the open-source software MuMax3, which solves the Landau-Lifshitz-Gilbert equation~\cite{Vansteenkiste2014}
\begin{eqnarray}
	\nonumber
	\frac{\mathrm{d}}{\mathrm{d}t} \vec{m} \left( \vec{r}, t \right)  = \frac{\gamma}{1 + \alpha^2} \, 
	\Big\{ \vec{m} \left( \vec{r}, t \right) \times \vec{B}_\text{eff} \left( \vec{r}, t \right) \\
	\label{eq:LLG}
	+ \alpha \left[ \vec{m} \left( \vec{r}, t \right) \times \left( \vec{m} \left( \vec{r}, t \right) \times \vec{B}_\text{eff} \left( \vec{r}, t \right) \right) \right] \Big\}
\end{eqnarray}
on a rectangular grid, we study spin-wave propagation in landscapes of the magnonic refractive index. Here, $\vec{m} = \vec{M} / M_\text{S}$ is the normalized vector of the magnetization~$\vec{M}$ with respect to the saturation magnetization~$M_\text{S}$. The effective magnetic field~$\vec{B}_\text{eff}$ includes the demagnetizing and exchange field, \mbox{Ruderman}-\mbox{Kittel}-\mbox{Kasuya}-\mbox{Yoshida} interaction, cubic and uniaxial anisotropy, thermal fluctuations, and the \mbox{Dzyaloshinskii}-\mbox{Moriya} interaction. $\gamma$~is the gyromagnetic ratio and~$\alpha$ describes the Gilbert damping parameter. To achieve spin-wave propagation over large distances, the sample system of choice for the simulation of spin-wave optics is a $\SI{5}{\micro\metre}$ thick yttrium iron garnet (YIG) film. YIG has the smallest known damping parameter~\cite{Chumak2017}~(here $\alpha = 10^{-4}$). Depending on the relative orientation of the external magnetic field with respect to the surface of the YIG film and the propagation direction, different spin-wave modes exist. If the external field is oriented out-of-plane, forward volume magnetostatic spin waves~(FVMSWs) exist. Due to symmetry reasons no in-plane propagation direction is favored~\cite{Damon1965}. Hence, the isofrequency curves are circles in analogy to the dispersion relation of light in conventional optics. In contrast, spin-wave propagation in in-plane magnetized samples is non-trivial: For spin-wave propagation parallel to the external field direction the modes are called backward volume magnetostatic spin waves~(BVMSWs). In perpendicular direction to the external field magnetostatic surface spin waves~(MSSWs) propagate. The resulting anisotropic isofrequency curves of BVMSW and MSSW modes are more complex. In first approximation, they are parabolic with orthogonal symmetry axis (see Refs.~\onlinecite{Vashkovsky2006} and~\onlinecite{Lock2008}).

\begin{figure}
	  \includegraphics{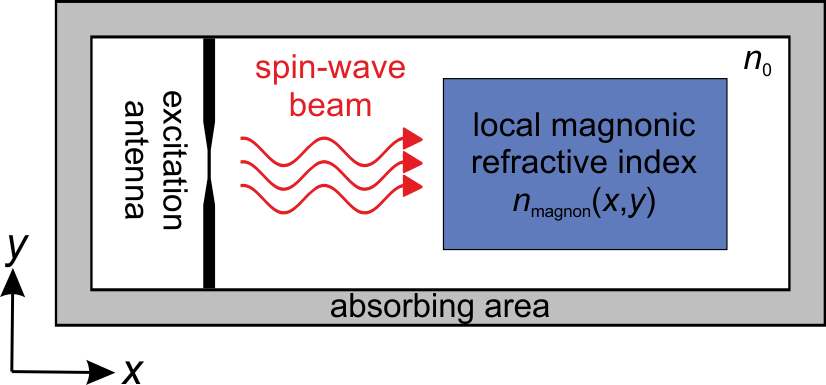}
		\caption{\label{fig:figure1} Scheme of the simulation area to excite spin-wave Gaussian beams and to study the propagation behavior in landscapes of the magnonic refractive index.}
\end{figure}

\begin{figure}
	  \includegraphics{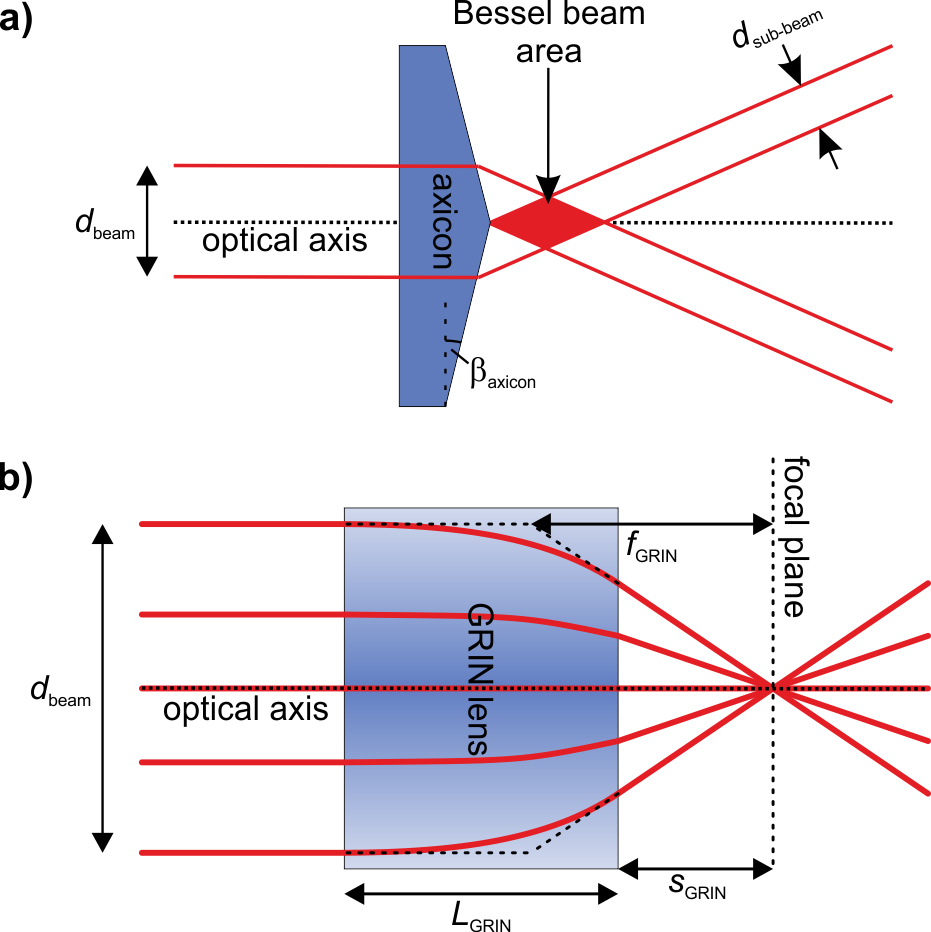}
		\caption{\label{fig:figure2} Scheme of two landscapes of the magnonic refractive index: \textbf{(a)} a spin-wave axicon is formed by a triangular shaped region with constant temperature, while \textbf{(b)} a GRIN lens contains a temperature gradient in a rectangular area.}
\end{figure}

The scheme of our calculations is shown in Figure~\ref{fig:figure1}.
The Oersted field of a narrowed microstrip antenna excites spin-wave Gaussian beams. To reduce undesired reflections at the edges of the simulation area we implement a $\SI{1}{\milli\metre}$ wide absorbing boundary in which the damping parameter increases exponentially to $\alpha = 1$. The blue marked region contains the magnonic refractive index profile that modifies the propagation of the spin-wave beam. We approximate the magnonic refractive index in analogy to optics as follows:
\begin{eqnarray}
n_\text{magnon} = \frac{c \, k}{\omega(k)} \text{ .}
\end{eqnarray}
Here, $c$ is the speed of light in vacuum, $k$ the spin-wave wavenumber, and $\omega(k)$ the dispersion relation obtained by solving the Landau-Lifshitz-Gilbert equation~\eqref{eq:LLG}. The dispersion relation of the respective spin-wave mode depends on many different parameters~\cite{Serga2010, Chumak2008a, Chumak2009_width, Chumak2009_current, Obry2013a, Gruszecki2018, Vogel2018}. Here, we change the local temperature~$T$ and therefore the saturation magnetization~$M_\text{S} \left( T \right)$~\cite{Vogel2015}: $M_\text{S} \propto -T$. It is assumed, that the sample is heated homogeneously across the film thickness. In this case $n_\text{magnon}$ does not depend on the $z$~coordinate and in the simulation only one cell unit is used for this coordinate. In the following, spin-wave propagation through landscapes of the magnonic refractive index equivalent to axicons and gradient-index (GRIN) lenses~(see Figure~\ref{fig:figure2}) is investigated.

\section{\label{sec:section1} Spin-wave axicons}
\begin{figure}
	  \includegraphics{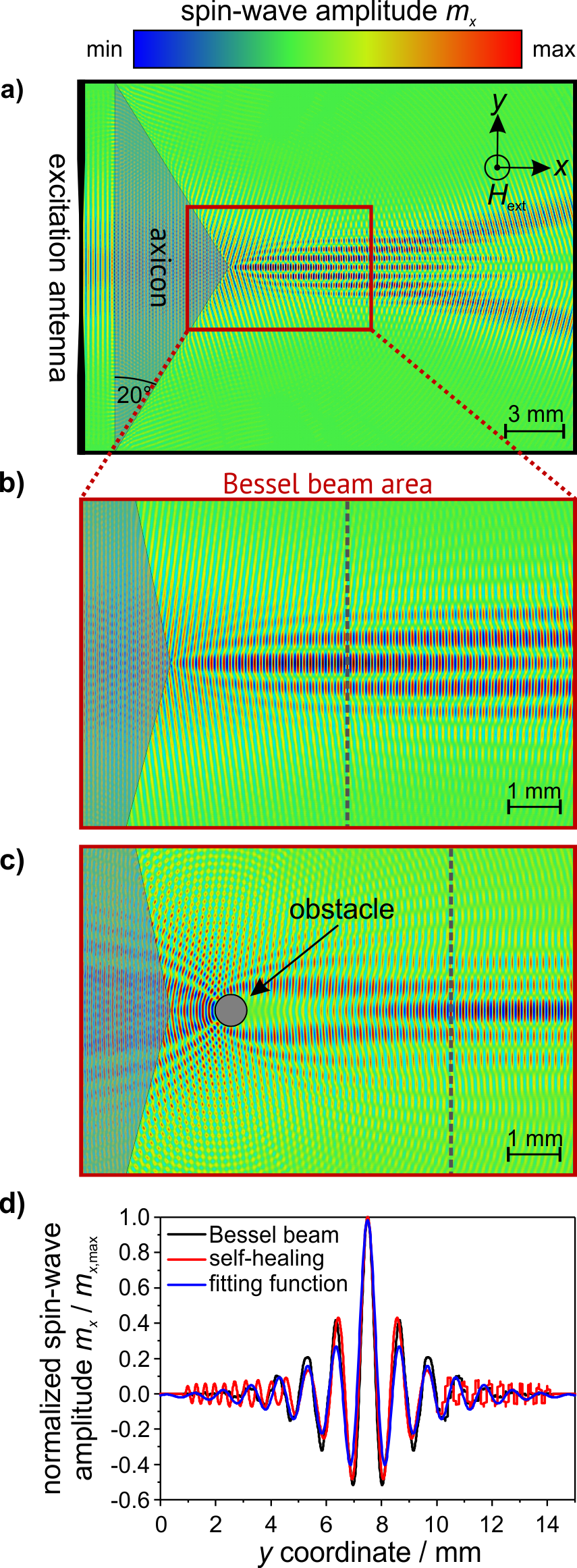}
		\caption{\label{fig:figure3} \textbf{a)} Simulation of a spin-wave Bessel beam. \textbf{b)} Enlarged Bessel beam area. \textbf{c)} Self-healing properties of Bessel beams after hitting a circular hole in the sample with a diameter of $\SI{500}{\micro\metre}$. \textbf{d)} Comparison of the normalized spin-wave amplitudes corresponding to the cross-sections in (b) and (c) (dotted lines).	The wavelength in the simulations is $\lambda = \SI{150}{\micro\metre}$. For an external magnetic field~$\mu_0 \, H_\mathrm{ext} = \SI{300}{\milli\tesla}$ this corresponds to a frequency~$f \approx \SI{3.78}{\giga\hertz}$. The characteristic angle~$\beta_\text{axicon}$ is $\ang{20}$ and the 30 x 15 $\text{mm}^2$ wide simulation area consists of 2048 x 2048 cells.}
\end{figure}

Besides the excitation of Gaussian beams, other beams with a more complex profile can be created such as a Bessel-Gaussian beam~\cite{Gori1987} which is used in photonics e.g. as optical tweezer~\cite{Arlt2001}. To realize such a beam in optics, a conically shaped lens (a so-called axicon) is commonly used (see Fig.~\ref{fig:figure2}a). Since spin waves propagate in thin films, a corresponding spin-wave axicon is the two-dimensional projection onto the film plane (an isosceles triangle). The incoming Gaussian beam with a diameter~$d_\text{beam}$ passes the optical element centered such that the beam is refracted at the backside in $+y$ and $-y$ direction. The initial beam is divided into two sub-beams with diameter~$d_\text{sub-beam} = d_\text{beam}/2$. Right next to the backside of the optical element both sub-beams interfere. In this area Bessel-Gaussian beams exist in good approximation. The size of the Bessel beam area depends on $d_\text{beam}$, the magnonic refractive index of the axicon and its surrounding area, as well as the characteristic angle~$\beta_\text{axicon}$. In Figure~\ref{fig:figure3} the obtained results of the micromagnetic simulations are shown.
The Bessel beam is non-diffractive~\cite{Durnin1987}. Moreover, the beam reconstructs its shape after impinging onto an obstacle. Hence, Bessel beams are self-healing~\cite{Garces2002}. The beam profile perpendicular to the propagation direction $x$ can be described by a Gaussian distribution and a Bessel function~$J_i$ of the first kind ($i$ is an integer). The normalized amplitude of the precessing magnetization's $x$ component~$m_x$ can be fitted using the equation
\begin{eqnarray}
\frac{m_x \left( y \right)}{m_{x\text{,max}}} = A \cdot J_0 \Big( k_\text{Bessel} \cdot \left( y - y_0 \right) \Big) \cdot \text{e}^{- \left( \frac{y - y_0}{w_\text{Bessel}} \right)^2} - B \text{ .}
\end{eqnarray}
In Fig.~\ref{fig:figure3}d beam profiles along cross-sections in the $y$ direction are plotted and compared to the obtained fitting function. The coefficient of determination $R^2$ is approximately 0.88. Thus, the observed beam is Bessel like. The parameters of the fitting procedure are listed in Table~\ref{tab:table1}.
\begin{table}
		\caption{\label{tab:table1} Parameter of the Bessel-Gaussian beam.}
		\begin{ruledtabular}
		\begin{tabular}{ll}
			  \textbf{parameter} & \textbf{fitted value} \\
			\hline
				$A$ & $\SI[separate-uncertainty]{0.995 \pm 0.008}{}$ \\
				$B$ & $\SI[separate-uncertainty]{0.011 \pm 0.002}{}$ \\
				$y_0$ & $\SI[separate-uncertainty]{7.500 \pm 0.002}{\milli\metre}$ \\
				$k_\mathrm{Bessel}$ & $\SI[separate-uncertainty]{6.138 \pm 0.009}{\per\milli\metre}$ \\
				$w_\mathrm{Bessel}$ & $\SI[separate-uncertainty]{4.290 \pm 0.118}{\milli\metre}$ \\
		\end{tabular}
		\end{ruledtabular}
\end{table}
Further simulations regarding BVMSW and MSSW modes are provided in the supplementary material (in-plane field $\mu_0 \, H_\mathrm{ext} = \SI{180}{\milli\tesla}$). The results qualitatively do not differ from Figure~\ref{fig:figure3}, although the corresponding dispersion relations are anisotropic.

\section{\label{sec:section2} Spin-wave GRIN lenses}
By continuously modifying the materials's parameters, and consequently the magnonic refractive index, the spin-wave propagation can be controlled further. In photonics, such continuous modifications are used in gradient-index optics~\cite{Saleh2008}. Typical examples are the Fata Morgana, GRIN fibers, and GRIN lenses. A GRIN lens is shown schematically in Fig.~\ref{fig:figure2}b.
\begin{figure*}
	  \includegraphics{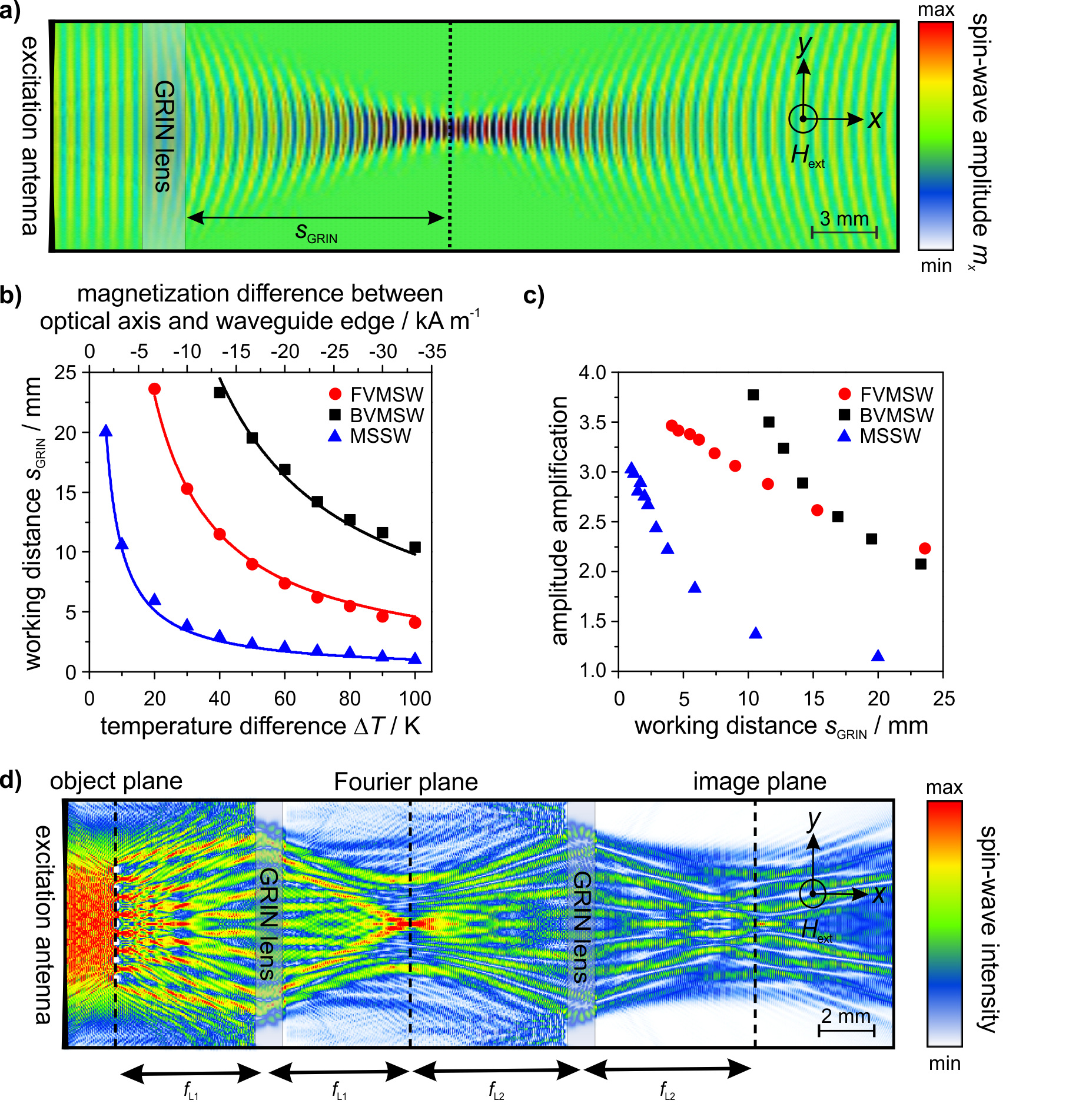}
		\caption{\label{fig:figure4} \textbf{a)} Focusing of spin waves using gradient-index lenses. Focusing can be achieved for all spin-wave modes. The external magnetic field is $\mu_0 \, H_\mathrm{ext} = \SI{300}{\milli\tesla}$ for FVMSWs and $\mu_0 \, H_\mathrm{ext} = \SI{180}{\milli\tesla}$ for BVMSWs and MSSWs. \textbf{b)} Working distances for different GRIN profiles dependent on the temperature difference between the optical axis and the edge of the lens. \textbf{c)} Amplitude amplification for the working distances shown in (b). \textbf{d)} GRIN lenses can be used to achieve spin-wave Fourier optics ($f_\mathrm{L1} \approx f_\mathrm{L2}$, $L_\mathrm{GRIN} = \SI{1}{\milli\metre}$, $\Delta T = \SI{50}{\kelvin}$).}
\end{figure*}
The rays are curved parabolically since the GRIN lens has a parabolic refractive index profile with the curvature~$g$~\cite{Saleh2008}:
\begin{eqnarray}
\label{eq:GRIN_n1} n \left( r \right) &&= n_0 - \Delta n \left( r \right) \\
\label{eq:GRIN_n2} &&= n_0 \cdot \left[ 1 - \frac{\Delta n \left( r \right)}{n_0} \right] \\
\label{eq:GRIN_n3} &&= n_0 \cdot \left[ 1 - \frac{g^2}{2} r^2 \right] \text{ .}
\end{eqnarray}
Here, $r$ is the radial distance to the optical axis where the refractive index is $n_0$. The focal distance~$f_\mathrm{GRIN}$ and working distance~$s_\mathrm{GRIN}$ are given by~($L_\mathrm{GRIN}$: thickness of the lens)~\cite{Saleh2008}:
\begin{eqnarray}
\label{eq:GRIN_f} f_\mathrm{GRIN} &= \frac{1}{n_0 \, g \, \sin \left( g \, L_\mathrm{GRIN} \right)} \approx \frac{1}{n_0 \, g^2 \, L_\mathrm{GRIN}} \text{ ,}\\
\label{eq:GRIN_s} s_\mathrm{GRIN} &= \frac{1}{n_0 \, g \, \tan \left( g \, L_\mathrm{GRIN} \right)} \approx \frac{1}{n_0 \, g^2 \, L_\mathrm{GRIN}} \text{ .}
\end{eqnarray}
For thin lenses $f_\mathrm{GRIN}$ and $s_\mathrm{GRIN}$ are equal ($g \, L_\mathrm{GRIN} \ll 1$). In the simulations we use parabolic temperature profiles to create gradient-index lenses:
\begin{eqnarray}
\label{eq:GRIN_T1} T \left( r \right) &&= T_0 + \Delta T \left( r \right) \\
\label{eq:GRIN_T2} &&= T_0 \cdot \left[ 1 + \frac{\Delta T \left( r \right)}{T_0} \right] \text{ .}
\end{eqnarray}
In Figure~\ref{fig:figure4} numerically obtained results for a thin lens with $L_\mathrm{GRIN} = \SI{2}{\milli\metre}$ and $\Delta T = \SI{40}{\kelvin}$ are shown. In the focal plane a pronounced increase of the spin-wave amplitude is visible. Comparing Eqs.~\eqref{eq:GRIN_n2} and~\eqref{eq:GRIN_n3} it follows:
\begin{eqnarray} \label{eq:GRIN_s_propto}
\Delta n \propto g^2 \, \stackrel[]{\eqref{eq:GRIN_s}}{\Longrightarrow} \, s_\mathrm{GRIN} \propto \frac{1}{\Delta n} \, \stackrel[]{\eqref{eq:GRIN_n2} \, \& \, \eqref{eq:GRIN_T2}}{\Longrightarrow} \, s_\mathrm{GRIN} \propto \frac{1}{\Delta T} \text{ .}
\end{eqnarray}
The working distance for different temperature or magnonic refractive index differences, respectively, is plotted in Fig.~\ref{fig:figure4}b. Additionally, the orientation of the external magnetic field is rotated in-plane to investigate the BVMSW and MSSW modes (see supplementary material, \mbox{$\mu_0 \, H_\mathrm{ext} = \SI{180}{\milli\tesla}$}). In case of BVMSWs the parabolic temperature profile has to be inverted, because the dispersion relation~$\omega \left( k \right)$ is monotonously decreasing with increasing $k$ (in contrast to FVMSWs and MSSWs, see Ref.~\onlinecite{Serga2010}). Equation~\eqref{eq:GRIN_s_propto} can be verified by fitting the obtained data in Fig.~\ref{fig:figure4}b.  The corresponding values of $R^2$ are larger than 0.97. By comparing the maximal amplitude in the focal plane with the value before passing the lens the amplification factor can be determined. This is shown in Fig.~\ref{fig:figure4}c for the investigated modes and the corresponding working distances from Fig.~\ref{fig:figure4}b. In the simulations, a maximal amplification factor of 3.8 for the amplitude and of 14.4 for the spin-wave intensity are observed.

\section{\label{sec:section3} Spin-wave Fourier optics}
In Fourier optics the propagation of light is described using Fourier analysis, and image modifications can be performed by filtering selected Fourier orders. Just as a lens with curved surfaces a GRIN lens can perform a Fourier transformation~\cite{Reino1987, Saleh2008}. Now we use such GRIN lenses to realize spin-wave Fourier optics (see Fig.~\ref{fig:figure4}d for the corresponding setup). The far field in the back focal plane of the first lens is the Fourier transform of the incoming field in the front focal plane ($2f$ setup). If the distance between the lenses equals the sum of the focal lengths, the arrangement is called a $4f$ setup. The object is positioned in the front focal plane of the first lens (object plane). The Fourier plane is in the back focal plane of the first lens or the front focal plane of the second lens, respectively. In the back focal plane of the second lens the image of the object is reconstructed (image plane). The magnification depends on the ratio of the lenses' focal distances $f_\mathrm{L1}$ and $f_\mathrm{L2}$. If $f_\mathrm{L1} = f_\mathrm{L2}$ the object is imaged without magnification. In Fig.~\ref{fig:figure4}d the object is a grating with 7 lines. Each line measures $\SI{100}{\micro\metre} \times \SI{200}{\micro\metre}$ and the lines are separated by $\SI{600}{\micro\metre}$. The corresponding wavevectors resulting from the Fourier transformation are visible as pronounced side maxima in the Fourier plane. The second GRIN lens reconstructs the grating as seven focal points in the image plane. The size of and the separation between the spots is approximately equal to the original grating, demonstrating a spin-wave hardware Fourier transformation.\\
\\
\textbf{Conclusions}\\
By changing the magnonic refractive index via e.g. modifying the local saturation magnetization the propagation of different spin-wave modes can be controlled. Despite very different and strongly anisotropic propagation characteristics, we report on axicons to realize Bessel-Gaussian beams as well as gradient-index lenses to focus spin waves in the FVMSW, BVMSW, and MSSW geometry. Furthermore, the properties of the GRIN lenses can be exploited to perform spin-wave Fourier optics. Since spin waves can be used for information processing~\cite{Karenowska2015}, magnonic gradient-index lenses have the potential to perform a Fourier transform with a single computational element in real-time.

\begin{acknowledgments}
Financial support by the DFG Transregional Collaborative Research Center \mbox{(SFB/TRR) 173} "\mbox{Spin + X} – Spin in its collective environment" within project B04 is gratefully acknowledged (DFG project number 268565370).
\end{acknowledgments}

\nocite{*}
\bibliography{./library}

\end{document}



\title[Spin-Wave Optical Elements: Towards Spin-Wave Fourier Optics -- SUPPLEMENTARY MATERIAL]{Spin-Wave Optical Elements: Towards Spin-Wave Fourier Optics\\SUPPLEMENTARY MATERIAL }

\author{M. Vogel}
\email{mvogel@physik.uni-kl.de}
\affiliation{Department of Physics and State Research Center OPTIMAS,  Technische Universität Kaiserslautern (TUK), Erwin-Schrödinger-Str. 56, 67663 Kaiserslautern, Germany}

\author{B. Hillebrands}
\affiliation{Department of Physics and State Research Center OPTIMAS,  Technische Universität Kaiserslautern (TUK), Erwin-Schrödinger-Str. 56, 67663 Kaiserslautern, Germany}

\author{G. von Freymann}
\affiliation{Fraunhofer-Institute for Industrial Mathematics ITWM, Fraunhofer-Platz 1, 67663 Kaiserslautern, Germany}
\affiliation{Department of Physics and State Research Center OPTIMAS,  Technische Universität Kaiserslautern (TUK), Erwin-Schrödinger-Str. 56, 67663 Kaiserslautern, Germany}

\date{\today}

\begin{abstract}
In the main text, axicons to create Bessel-Gaussian beams are discussed in the FMVSW geometry. The external magnetic field is aligned out-of-plane and the thickness of the yttrium iron garnet film is $\SI{5}{\micro\metre}$. Despite the strong anisotropy in in-plane magnetized films it is also possible to excite such beams in the BVMSW and MSSW geometry. In Figure~\ref{fig:figure_S1} the corresponding results of micromagnetic simulations are shown. In contrast to FVMSW modes, the phase fronts of the sub-beams are not perpendicular to the propagation direction. Moreover, gradient-index lenses are also possible in in-plane magnetized samples~(see Figure~\ref{fig:figure_S2}). The obtained results in Figure~\ref{fig:figure_S1} and Figure~\ref{fig:figure_S2} qualitatively do not differ from the FVMSW configuration shown in the main text.
\end{abstract}

\maketitle

\begin{figure}
	  \includegraphics{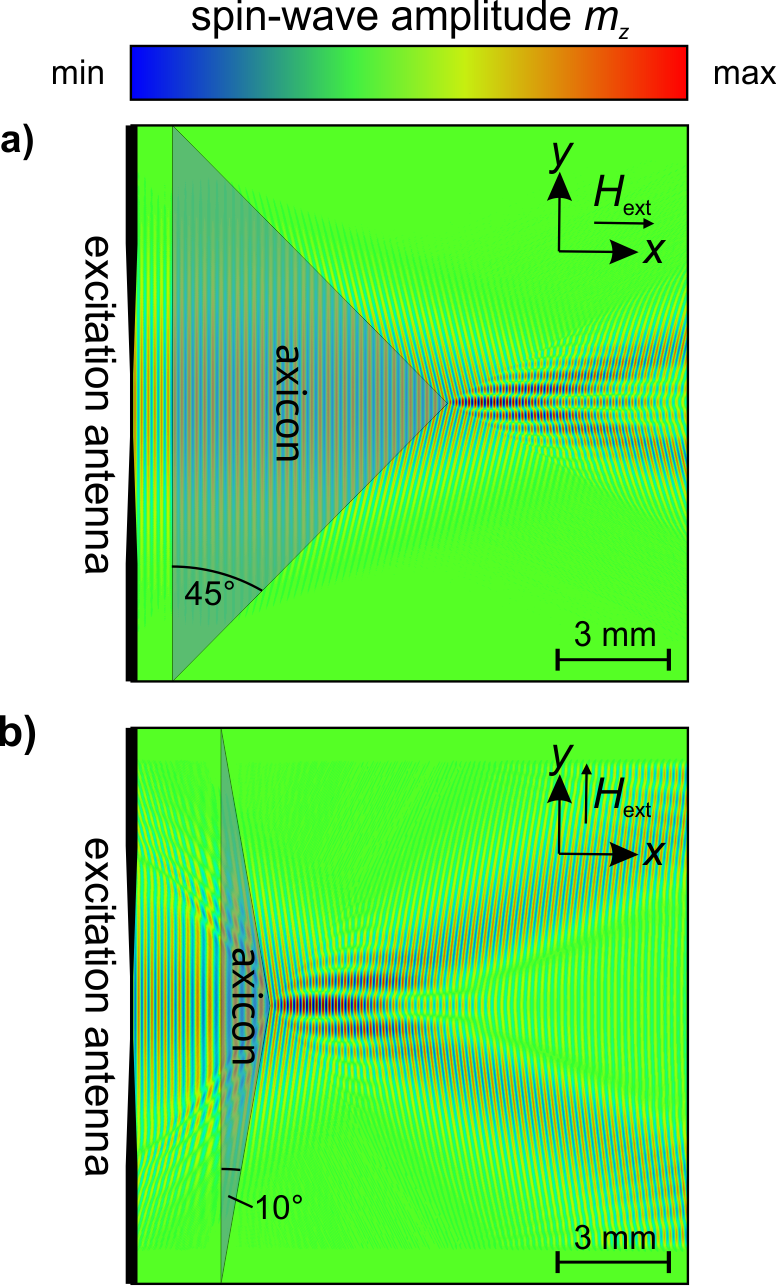}
		\caption{\label{fig:figure_S1} Axicons to realize Bessel-Gaussian beams for BVMSW \textbf{(a)} and MSSW modes \textbf{(b)}. The external magnetic field is aligned in-plane ($\mu_0 H_\text{ext} = \SI{180}{\milli\tesla}$). The background temperature is $\SI{298}{\kelvin}$. The axicon's temperature differs by $\SI{10}{\kelvin}$.}
\end{figure}
\begin{figure}
	  \includegraphics{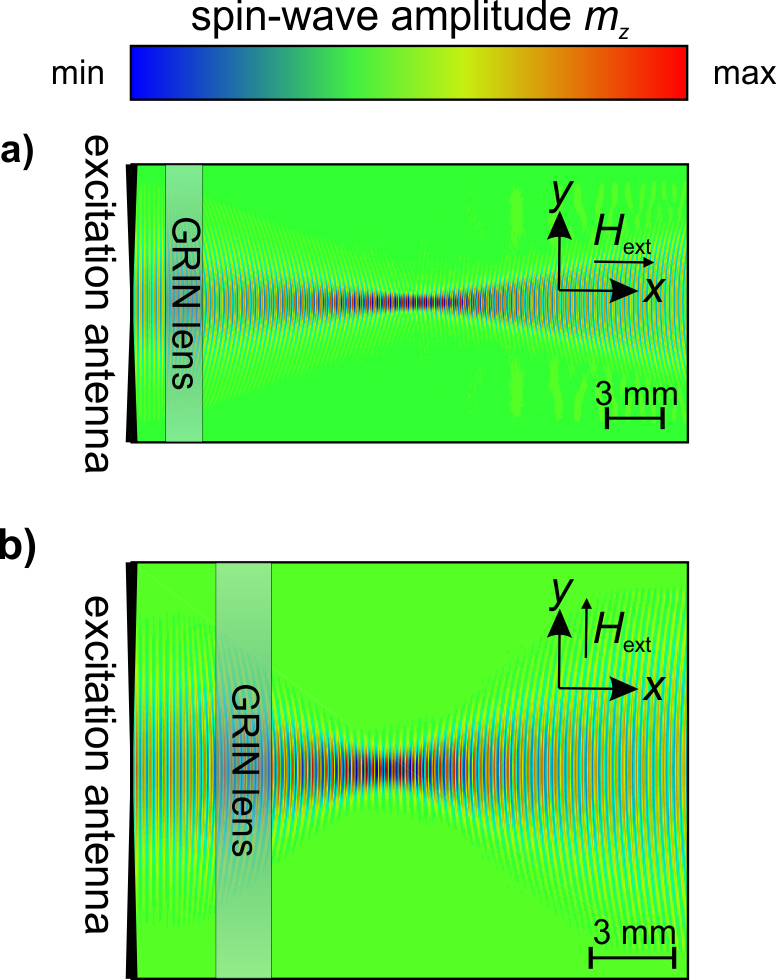}
		\caption{\label{fig:figure_S2} GRIN lenses to focus BVMSWs \textbf{(a)} and MSSWs \textbf{(b)}. The external magnetic field is aligned in-plane. The magnitude is $\mu_0 H_\text{ext} = \SI{180}{\milli\tesla}$. The shown lenses are $\SI{2}{\milli\metre}$ wide and the temperature difference~$\Delta T$ is $\SI{40}{\kelvin}$ in (a) and $\SI{10}{\kelvin}$ in (b)}
\end{figure}